\documentclass[12pt]{article} 
\usepackage{color}
\usepackage{amsfonts}
\usepackage{latexsym}
\usepackage{amsmath,amssymb}
\usepackage{verbatim}
\usepackage{datetime,cite}

\newcommand{\bea}{\begin{eqnarray}}
\newcommand{\eea}{\end{eqnarray}}
\newcommand{\be}{\begin{equation}}
\newcommand{\ee}{\end{equation}}
\newcommand{\ba}{\begin{array}}
\newcommand{\ea}{\end{array}}

\def\nn{\nonumber}
\def\half{{1\over2}}
\def\p{\partial}
\def\eps{\epsilon}

\numberwithin{equation}{section}

\def\cT{\mathcal{T}}
\def\cL{\mathcal{L}}
\def\cJ{\mathcal{P}}
\def\cM{\mathcal{M}}
\def\tr{\hbox{Tr}}
\def\sdelta{\slash\hspace{-6pt}\delta}

\definecolor{Wei}{rgb}{0.65,0.0,0}
\definecolor{Geo}{rgb}{0.1,0,0.75}

\begin{document}

\begin{center}\vspace{2cm}
{ \LARGE {{New Boundary Conditions for AdS$_3$}}}

\vspace{1cm}

Geoffrey Comp\`ere$^{*\diamond}$, Wei Song$^*$ and Andrew Strominger$^{*\dagger}$

\vspace{0.8cm}

{\it  $*$Center for the Fundamental Laws of Nature, Harvard University,\\
Cambridge, MA, USA}%\\\vspace{0.6cm}

{\it  $\diamond$ Physique Th\'eorique et Math\'ematique, Universit\'e Libre de Bruxelles,\\
 Bruxelles, Belgium}%\\\vspace{0.6cm}

{\it  $^\dagger$Radcliffe Institute for Advanced Study, %Harvard University,\\
Cambridge, MA, USA}

\vspace{0.5cm}
%\today

\vspace{0.5cm}

\vspace{1.0cm}

\end{center}

\begin{abstract}

New chiral boundary conditions are found for quantum gravity with matter on AdS$_3$.  The associated asymptotic symmetry group is generated by a single right-moving $U(1)$ Kac-Moody-Virasoro algebra with $c_R={3 \ell \over 2 G}$. The Kac-Moody zero mode generates global left-moving translations and equals, for a BTZ black hole,  the  sum of the total mass and spin.  The level is positive about the global vacuum and  negative in the black hole sector, corresponding to  ergosphere formation. Realizations arising in Chern-Simons gravity and  string theory  are analyzed. The new boundary conditions are shown to naturally arise for warped AdS$_3$ in the limit that the warp parameter is taken to zero.

\vspace{1cm}

\end{abstract}
\thispagestyle{empty}

\pagebreak
\setcounter{tocdepth}{2}

\tableofcontents

\section{Introduction}

In a seminal paper \cite{Brown:1986nw} Brown and Henneaux found consistent boundary conditions for quantum gravity (plus matter) on AdS$_3$  and showed that they imply the semiclassical states  form representations of left and right Virasoro algebras with $c_{L,R}={3\ell \over 2 G}$, where $G$ is Newton's constant and $\ell$ is the AdS$_3$ radius.  The Virasoro symmetries act infinitesimaly on the coordinates $t^\pm$ of the boundary cylinder as 
\be \label{dsx} \delta t^+= \epsilon^+(t^+),~~~\delta t^-=\epsilon^-(t^-).\ee
 Importantly, Brown-Henneaux boundary conditions admit all BTZ black holes  \cite{Banados:1992wn,Banados:1992gq}.  Generalizations and modifications are analyzed in $e.g.$ \cite{%Polyakov:1987zb,Skenderis:1999nb,
 Saida:1999ec,Henneaux:2002wm,Hotta:2008yq,Compere:2008us,Henneaux:2009pw,Liu:2009kc,Henneaux:2010fy}.

The analysis of \cite{Brown:1986nw} considers only  parity-invariant boundary conditions.   In this paper we drop this restriction and find a new set of non-chiral but self-consistent boundary conditions for AdS$_3$ quantum  gravity (plus matter) admitting BTZ black holes.
The asymptotic symmetry generators form  a single right-moving $c_R={3\ell \over 2 G}$ noncompact $U(1)$ Kac-Moody-Virasoro algebra and (\ref{dsx}) is replaced by 
\be \label{dsx2} \delta t^+= \epsilon(t^+),~~~\delta t^-=\sigma(t^+).\ee
Note that the diffeomorphisms of $both$ $t^-$ and $t^+$ depend only on $t^+$.  
The zero mode $\delta t^-={\rm constant}$ of the Kac-Moody symmetry is a global left translation.

We further compute the (nonzero but normalization-dependent) Kac-Moody level. A salient feature is that it is positive around the global vacuum, but negative in the black hole sector. This negative level is related to fact that the vector fields generating the Kac-Moody symmetry are spacelike in the black hole sector. This can be thought of as a type of ergosphere and entails negative energy modes and superradiant-type  instabilities of the black holes. These instabilities are a simple versions of the ones encountered in a variety of contexts including AdS$_2$ fragmentation \cite{Maldacena:1998uz}  and Kerr superradiance. We hope that the current work will provide a simple useful context in which to study them.

The discovery of these new boundary conditions was sparked by a circle of recent investigations \cite{Anninos:2008fx,Compere:2008cv,Guica:2008mu,Compere:2009zj,Castro:2009jf, Compere:2009qm,Anninos:2010pm,Guica:2010sw,Hofman:2011zj,ElShowk:2011cm,Song:2011sr,Guica:2011ia,Azeyanagi:2012zd,Detournay:2012pc} involving warped AdS$_3$ spacetimes, near-horizon extreme Kerr,  string compactifications and their potential dual warped CFT$_2$s - all of which have in common global $SL(2,\mathbb{R})\times U(1)$ symmetries.  Asymptotic boundary conditions have been discussed in these contexts \cite{Compere:2008cv,Guica:2008mu,Compere:2009zj,Loran:2009cr,
Azeyanagi:2009wf,Balasubramanian:2009bg,Blagojevic:2009ek,Anninos:2010pm,ElShowk:2011cm,Henneaux:2011hv,Song:2011sr}
%Do we cite ?
which contain one left Virasoro or one right Kac-Moody Virasoro but none with both a simultaneous left and right Virasoro. In contemplating this puzzle we realized that if consistent right Kac-Moody Virasoro boundary conditions exist  for warped AdS$_3$,  they should persist in the limit
that the warping is deformed to zero and ordinary AdS$_3$ reappears. We find this is indeed the case: we spell out the new AdS$_3$ boundary conditions, verify their consistency, compute the central extensions and work out several examples.

It is important to note that the semiclassical consistency of the chiral boundary conditions found here does not guarantee the existence of any non-trivial fully quantum theory obeying  these boundary conditions. So far no clear example is known. Nevertheless it appears to us that  the semiclassical self-consistency is highly non-trivial and motivates the search for such quantum theories. A potential warped stringy example is discussed in \cite{Azeyanagi:2012zd}, and one may attempt to impose the new chiral boundary conditions on any string compactification to AdS$_3$.  Other candidates are supplied by the chiral AdS$_3$ Chern-Simon gravity solved semiclassically in section 3, and its holographic Liouville dual analyzed in \cite{css2}. The quantum versions of these theories remains to be studied.  

In a companion paper \cite{css2}, we consider  the problem from a dual purely two-dimensional  perspective of (classical) Liouville gravity. Chiral gauge conditions for the two-dimensional metric are found for which the residual symmetry is generated by a single right-moving  Virasoro-Kac-Moody, as opposed to the familiar two Virasoros found in conformal gauge. The action, equations of motion and Dirac brackets of the resulting nonlinear chiral theory of Liouville gravity are identical to those of the AdS$_3$ Chern-Simons gravity with new boundary conditions constructed in this paper.  Hence two-dimensional chiral Liouville theory and AdS$_3$ Chern-Simons gravity with the new boundary conditions are equivalent.

The paper is organized as follows. In section 2 we specify the new boundary conditions, show that the charges are finite and integrable and derive the chiral Kac-Moody-Virasoro asymptotic symmetry algebra including the central extensions.
In section 3 we construct the simplest example by applying the boundary conditions to the $SL(2,\mathbb{R})_R\times SL(2,\mathbb{R})_L$ Chern-Simons formulation of pure Einstein gravity on AdS$_3$. The application of Brown-Henneaux boundary conditions is well-known \cite{Coussaert:1995zp} to yield a single non-chiral Liouville scalar with
central charges $c_{R,L}={3\ell \over 2 G}$. The new boundary conditions yield instead two right-moving chiral scalars with a single  $c_{R}={3\ell \over 2 G}$ Virasoro-Kac-Moody, which together form a chiral version of Liouville gravity \cite{css2}. The  Kac-Moody level is also computed as a function of the current normalization. In section 4 the realization of the new  boundary conditions in weak-field perturbation theory is presented. We perturbatively construct the lowest weight representations for a scalar field coupled to gravity. The lowest weight state is a single scalar particle in the center of AdS$_3$. The descendants are arbitrary excited single scalar particle states surrounded by a gas with arbitrary numbers of right-moving boundary gravitons and right-moving boundary photons.
In section 5 we discuss a  compactification of string theory to warped AdS$_3$ which has been analyzed in several papers \cite{Detournay:2010rh,ElShowk:2011cm,Song:2011sr,Azeyanagi:2012zd}. Consistent boundary conditions are given for general warp parameter, and then shown to reduce precisely to
the new chiral boundary conditions in the limit that the warp parameter is taken to zero.

\section{Asymptotic symmetry group analysis}
In this section we present the new boundary conditions and derive the asymptotic symmetry algebra.

\subsection{New boundary conditions}

Let $t^\pm = t \pm \phi$ where $\phi \sim \phi +2\pi$. We impose the following asymptotically AdS$_3$ boundary conditions on the $3d$ metric

\bea
g_{rr} &=& \frac{\ell^2 }{r^2}+O(r^{-4}),\nn\\
g_{r\pm} &=& O(r^{-3}), \nn\\
g_{+-} &=&- \frac{\ell^2 r^2}{2}+O(r^0),\label{newBC}\\
g_{++} &=& \p_+\bar P(t^+)\ell^2  r^2+O(r^0),\nn\\
g_{--} &=& 4G\ell  \Delta +O(r^{-1}),\nn
\eea
where $\Delta$ is a fixed constant and $\p_+\bar P(t^+)$ is periodic. In Fefferman-Graham coordinates, any three-dimensional Einstein metric admits the expansion
\bea
ds^2 = \ell^2 \frac{dr^2}{r^2}+\ell^2  r^2 \left( g^{(0)}_{ab} + \frac{1}{r^2}g^{(2)}_{ab} +O(r^{-3}) \right) dx^a dx^b\, .\label{FG2}
\eea
(\ref{newBC}) then reduces to
\bea
&&g^{(0)}_{--} = 0,\qquad  g_{++}^{(0)} = \p_+\bar P(t^+),\qquad  g^{(0)}_{-+}=g^{(0)}_{+-}= -\frac{1}{2},\nn\\
&&g^{(2)}_{--}=\frac{4G}{\ell} \Delta  ,\label{BCW}
\eea with subleading terms unconstrained by the boundary conditions.
This can be compared to the standard Brown-Henneaux boundary conditions \cite{Brown:1986nw} in Fefferman-Graham coordinates
\bea
g^{(0)}_{--} = g_{++}^{(0)}=0,\qquad  &g^{(0)}_{-+}=g^{(0)}_{+-}=- \frac{1}{2}  ,\label{BCWBH}
%g^{(2)}_{++}=g^{(2)}_{++}(t^+),\qquad &g^{(2)}_{--}=g^{(2)}_{--}(t^-)  .
\eea
with subleading terms unconstrained. The new boundary conditions differ in several respects from these. They are chiral since they do not treat left and right movers symmetrically. The leading order part of the metric $g_{++}^{(0)}$ is allowed to fluctuate with right-movers while the subleading part of the metric $g_{--}^{(2)}$ is restricted to be a fixed constant. In both cases, the boundary metric $g_{ab}^{(0)}$ is restricted to be Ricci-flat, $R_{(0)} = 0$.

Note that an alternative version of the boundary conditions exists where $\Delta$ is allowed to be varied while $\bar P(t^+)$ is periodic. We discuss these alternative boundary conditions separately in Appendix B.

% {\color{Geo} [XX]} The natural time coordinate which is independent from fluctuating fields is $t = (t^++t^-)/2$. Its norm is asymptotically
%\bea
%|\p_t|^2 = -l^2 r^2 (1-\p_+\bar P(t^+))+O(r^0),
%\eea
%which could become positive. This means that there is effectively an ergoregion in the asymptotic region. Also, the norm of $\p_\phi$ could become negative along part of the circle $\phi$ but the function $T \equiv (t^++t^--\bar P(t^+))/2$ provides with a global time function in the asymptotic region, $g^{\mu\nu}\p_\mu T \p_\nu T <0$. Causality in the bulk is therefore ensured as it should, since $\bar P(t^+)$ can be merely generated by the diffeomorphism $t^- \rightarrow t^- - \bar P(t^+)$.

\subsection{Variational principle}

We consider gravitational theories whose metrics  are asymptotically governed by the vacuum Einstein equation. After the addition of the standard Gibbons-Hawking term and counterterms, the variation of the action is given by
\bea
\delta S_0 = \frac{\ell}{16\pi G  }\int d^2x \sqrt{-g_{(0)}}\, g_{(2)}^{ab}\, \delta g_{(0)ab}\label{varS}
\eea
where indices are raised with the boundary metric $g^{(0)ab}$. The total action is defined as $S_0$ complemented by a chiral boundary term, 
\bea
S = S_0 + \frac{\Delta}{4\pi}\int dt^+ dt^- \sqrt{-g_{(0)}}g_{(0)}^{--}.
\eea
With the boundary conditions \eqref{newBC}, the variation of the action (\ref{varS}) is given by
\bea
\delta S =- \frac{1}{2\pi}\int dt^+ dt^- \p_+\bar P \delta \Delta =0.
\eea

\subsection{Nonlinear solutions}

For pure 3D Einstein gravity with no matter, the Fefferman-Graham expansion terminates after fourth order \cite{Skenderis:1999nb} :
\bea
ds^2 = \ell^2  \frac{dr^2}{r^2}+\ell^2 r^2 (g_{(0)ab}+\frac{g_{(2)ab}}{r^2} + \frac{g_{(4)ab}}{r^4})dx^a dx^b\label{eq:m}
\eea
where $g_{(4)ab} = \frac{1}{4}g_{(2)ac} g^{(0)cd} g_{(2)db}$ and $g_{(2)ab}$ is constrained by the equations of motion.
It is then a straightforward exercise to solve Einstein's equations with the boundary conditions \eqref{newBC}.
The general solution obeying the boundary conditions can be written as
\bea
ds^2 &&= \frac{\ell^2 }{r^2}dr^2-\ell^2  r^2 dt^+\big( dt^- - \p_+\bar P(t^+)dt^+\big)\nonumber \\ &&
+ 4G\ell  \Big[ \bar L(t^+)(dt^+)^2 +\Delta \big(dt^--\p_+\bar P(t^+)dt^+\big)^2 \Big] \nonumber \\
&& -\frac{16 G^2 \Delta}{r^2} \bar L(t^+) dt^+ \big(dt^--\p_+\bar P(t^+) dt^+ \big). \label{dressedmetric}
\eea
In terms of Fefferman-Graham coefficients, one has
\bea
&&g^{(0)}_{--} = 0,\qquad g^{(0)}_{+-} =- \half,\qquad g^{(0)}_{++} = \p_+\bar P, \label{metricuptosecondorder}  \\
&&g^{(2)}_{--}=\frac{\Delta}{k}\qquad g^{(2)}_{+-} = -\frac{\Delta}{k} \, \p_+\bar P,\qquad
g^{(2)}_{++} = \frac{1}{k}  (\bar L(t^+)+ \Delta\, (\p_+\bar P)^2),\nn
\eea
where we defined for convenience
\bea
k\equiv \frac{\ell}{4G}.
\eea
In the special case of vanishing $\p_+\bar P$ and $\bar L(t^+)=\bar \Delta$ the solution becomes
\be
\frac{ds^2}{\ell^2} = \frac{dr^2}{r^2}-  r^2 dt^+ dt^- + \frac{\bar \Delta}{k}
(dt^+)^2  + \frac{\Delta}{k} (dt^-)^2 -\frac{\Delta\bar \Delta}{k^2 r^2} dt^+  dt^-,
\ee
which is just the BTZ black hole with $\ell M=\Delta+\bar \Delta$ and $J=\Delta-\bar \Delta$.
As we will see below the general solution can be interpreted as a BTZ  black hole dressed by gases of boundary gravitons and photons represented by the non-zero modes of $\bar L(t^+)$ and $\bar P(t^+)$. % {\color{Geo} [XX]} Some properties of the dressed BTZ black holes are similar to the ones of the original BTZ. In particular, the generator of the horizon
%\bea
%\xi = (1+\Omega)\p_+ + (1-\Omega + (1+\Omega)\p_+\bar P)\p_-
%\eea
%where $\Omega$ is the angular velocity is non-spacelike outside the black hole, which prevents the existence of a bulk ergoregion. Nevertheless, the presence of boundary photons allow $\p_t$ to become spacelike which signals a ``boundary ergoregion'', as mentioned earlier.

Of course when matter sources are present, the solutions are modified in the interior. In the following we assume that the matter sources fall off sufficiently rapidly so that (\ref{FG2})-(\ref{metricuptosecondorder}) can be used in the asymptotic analysis.

\subsection{Asymptotic symmetry algebra}
\label{sec:chargesmetric}

This phase space of metrics is preserved under the action of the asymptotic symmetry algebra. This consists of a right-moving Virasoro algebra and a ``crossover" right-moving $U(1)$ current algebra whose zero mode is the left-moving generator $\p_-$, and is generated by
\bea
\xi_R(\epsilon) &=& \eps(t^+)\p_+ - \frac{r}{2}\eps^\prime(t^+)\p_r + (subleading),\\
\label{crs} \eta(\sigma) &=& \sigma(t^+)\p_-+ (subleading).
\eea
Note that asymptotically the vector field $\eta$ goes from spacelike to timelike when $\Delta$ crosses zero.  We will see in the following that this implies a sign change in the level of the associated  current algebra.

The canonical infinitesimal charges associated with these generators  as defined in \cite{Abbott:1981ff,Barnich:2001jy,Barnich:2007bf} can be directly integrated on the phase space. They are given by
\bea
Q_{\xi_R} &=&  \frac{1}{2\pi}\int_0^{2\pi} d\phi\; \eps(t^+) \left( \bar L(t^+)  -\Delta(\p_+\bar P(t^+))^2 \right),\label{ch1}\\
Q_{\eta} &=& \frac{1}{2\pi}\int_0^{2\pi} d\phi\; \sigma(t^+) (\Delta+ 2\Delta \p_+\bar P(t^+)).\label{ch2}
\eea
The charges are finite and conserved and determined up to a constant background value. Here, we set the zero mode of the charge $Q_\eta$ to $\Delta$ in order to reproduce the usual charge when $\Delta$ is allowed to vary. When $\p_+\bar P =\delta \p_+\bar P=0$,  $\p_+$ is canonically associated with $\bar \Delta$ and $\p_-$ with $\Delta$.  For the BTZ black hole, the energy is $M= Q_{\p_t} = \frac{1}{\ell}(Q_{\p_+}+Q_{\p_-}) =\frac{ \bar \Delta + \Delta}{\ell}$ and the angular momentum is $J=Q_{-\p_\phi} = -Q_{\p_+}+Q_{\p_-}=\Delta - \bar \Delta$.

Setting $\epsilon =e^{int^+}$, $\sigma = e^{i n t^+}$ we define $\bar \cL_n = Q_{\xi_R}$, $\bar \cJ_n = Q_{\eta}$. The Dirac bracket between these generators is given by
\bea
i \{\bar \cL_m,\bar \cL_n\} &=& (m-n)\bar \cL_{m+n}+\frac{c_R}{12}m^3\delta_{m,-n},\\
i\{\bar \cL_m,\bar \cJ_n\}&=&-n \bar \cJ_{m+n},\\
i\{\bar \cJ_m,\bar \cJ_n\} &=& \frac{k_{KM}}{2} m \delta_{m,-n}.
\eea
The central charge and level of the current algebra are given by
\bea
c_R = \frac{3\ell }{2G},\qquad k_{KM} = -4 \Delta \, .\label{cc}
\eea

The Kac-Moody level $k_{KM} = k$ is positive near the AdS$_3$ vacuum $\Delta = -k/4$ but becomes negative
in the presence of a $\Delta>0$ black hole. At the classical level, this means that boundary  photons lower the right moving energy ($\cL_0$) of the black hole, potentially indicating an instability. 
This is closely related to the facts that for nonzero $\p_+\bar P$ the vector field $\p_t$ is in general no  longer everywhere timelike  and  $\eta(\sigma)$ is spacelike. This region may be thought of as an ergosphere.   In this regard AdS$_3$ with the new boundary conditions may provide useful insight into the more complicated Kerr ergosphere and associated superradiant instabilities. We hope to understand this better and expect the recent analysis of \cite{Detournay:2012pc} may be relevant.

Defining the boson $\psi$ by
\be \label{ps} \psi(t^+,t^-)=  2\Delta (- t^- +\bar P(t^+)),\ee
one finds
\be
\bar \cJ_n =
\frac{1}{2\pi}\int_0^{2\pi} d\phi e^{i n t^+} \p_+\psi + \delta_n^0 \Delta,
\ee
and
\be
\bar \cL_n  =%\bar{\mathcal{L}}_b+
 \frac{1}{2\pi}\int_0^{2\pi} d\phi e^{i n t^+} (\bar L(t^+) +{1 \over k_{KM}}(\p_+\psi)^2),
\ee
which is the standard Sugawara formula for the stress tensor.

%%%%%%%%%%%%%%%%
\section{AdS$_3$ Chern-Simons gravity}
In this section we derive the dual warped CFT$_2$ associated with the new boundary conditions to pure gravity on AdS$_3$ in the Chern-Simons formulation and verify that it contains the Virasoro-Kac-Moody structure predicted by the asymptotic symmetry analysis. We follow the well-known analysis for the Brown-Henneaux case \cite{Coussaert:1995zp}, which was the first construction of a CFT$_2$ with $c_{R,L}={3\ell \over 2G}$. For simplicity we restrict here to the classical (large level) limit. 

  We note that pure 3D Einstein gravity appears unlikely to be a fully consistent nonperturbative quantum theory of gravity with any boundary conditions because, among other reasons, it does not have enough degrees of freedom to account for black hole entropy.  %{\color{Geo} [XX]} \cite{Maloney:2007ud}.
  Nevertheless it does appear to model a subsector of more complete gravitational theories and the analysis here gives insight into the structure of our boundary conditions.

\subsection{Chern-Simons formalism}
\label{CSform}
Three dimensional Einstein gravity with a negative cosmological constant can be formulated as $SL(2,\mathbb{R})_L\times SL(2,\mathbb{R})_R$ Chern-Simons gauge theory, with the action \cite{Achucarro:1987vz,Witten:1988hc}
\bea
S_E[A,\bar{A}]=S_{k}[A]+S_{-k}[\bar{A}]\label{SE}
\eea
where
\bea
S_{k}[A]&=&{k\over4\pi}\int_{\cM}  \tr (A \wedge dA+{2\over3} A\wedge A\wedge A) - {k\over4\pi}\int_{\Sigma} d^2x \tr (A_tA_\phi)\nn\\
&=&-{k\over4\pi}\int_{\cM} d^3x \tr (- \dot{A}_\phi A_r+\dot{A}_rA_\phi-2A_tF_{\phi r})
\eea
and \be k={\ell\over4G}.\ee
 Here $\Sigma$ is the boundary of $\cM$ at spatial infinity. The equations of motion are given by \be F \equiv dA+A \wedge A=0,\quad  \bar F \equiv d\bar{A}+\bar{A} \wedge \bar{A}=0\ee
where \bea
A&=&\omega_\alpha\,^\beta+\frac{1}{\ell}e_\alpha\,^\beta ,\qquad
\bar{A}=\omega_\alpha\,^\beta-\frac{1}{\ell}e_\alpha\,^\beta .
\eea
Here and elsewhere unbarred quantities refer to the left sector, and barred quantities to the right sector. More details on our conventions can be found in Appendix A.

\subsection{The new boundary conditions}

In terms of triads the new $AdS_3$ boundary conditions (\ref{newBC}) in Fefferman-Graham gauge become
\bea
\frac{e^{(+)}}{\ell}&=& r (dt^-- \p_+\bar P(t^+)dt^+)-\frac{1}{kr}\bar L(t^+)dt^+ +O(r^{-2})dt^\pm +O(r^{-4})dr,\nonumber\\
\frac{e^{(-)}}{\ell}&=&rdt^+-\frac{ \Delta}{k r}(dt^--\p_+\bar P(t^+)dt^+)+O(r^{-2})dt^\pm +O(r^{-4})dr,\nonumber\\
\frac{e^{(3)}}{\ell}&=&{dr\over r}+O(r^{-2})dt^\pm +O(r^{-4})dr,
\eea
where we specified the fluctuating mode $\bar L(t^+)$ implied by the asymptotic equations of motion. In terms of the gauge fields this is
\bea
A&=&\left( \begin{array}{cc}
  {dr\over2 r}&\frac{\Delta}{k r}(dt^--\p_+\bar P(t^+)dt^+) \\
   r (dt^--\p_+\bar P(t^+)dt^+)  & -{dr\over2 r}\end{array}
\right)\label{BC_A1} \nonumber\\
&& \quad+\left( \begin{array}{cc}
  O(r^{-2})& O (r^{-2})  \\
 O(r^{-1})  & O(r^{-2}) \end{array}
\right),\\
\bar{A}&=&\left( \begin{array}{cc}
  -{dr\over 2r}+O (r^{-2}) &  r dt^+ +{O (r^{-2})} \\
  \frac{1}{k r}\bar L(t^+)dt^++O(r^{-2})  & {dr\over 2r}+O (r^{-2}) \end{array}
\right). \label{BC_A2}
\eea
The boundary conditions on the (barred) right moving  part are the same as Brown-Henneaux \cite{Brown:1986nw}, or  Coussaert-Henneaux-van Driel \cite{Coussaert:1995zp} in the Chern-Simons formalism. The (unbarred) left moving sector is different both at leading order and at subleading order. There is a crossover right moving Kac-Moody current at leading order. Since $\Delta$ is constant, there is also a constraint on the subleading piece of $ A_{(+)}$, which prevents the excitation of the left-moving Virasoro modes. The right-moving Virasoro modes are still present in the subleading part of $\bar {A}_{(-)}$.
The new boundary conditions can be obtained from the Brown-Henneaux ones by simultaneously applying the change of coordinates
$t^-  \rightarrow t^ - - \bar P(t^+)$ and fixing $L(t^-) = \Delta $.

\subsection{Conserved charges}
The infinitesimal conserved charges associated with the gauge parameters $\Lambda$ and $\bar\Lambda$ defined by
\bea
\delta A_\mu = D_\mu \Lambda = \p_\mu \Lambda + [A_\mu,\Lambda], \qquad \delta \bar A_\mu = \bar D_\mu \bar \Lambda = \p_\mu \bar \Lambda + [\bar A_\mu,\bar \Lambda]
\eea
are given by
\bea
\sdelta Q_\xi = \frac{k}{2\pi}\int_{2\pi}^{0} d\phi\; \hbox{Tr} \left( \delta A_\phi \Lambda - \delta \bar A_\phi \bar \Lambda \right).
\eea
Up to an irrelevant local Lorentz transformation, a diffeomorphism is given by $\Lambda = \xi^\mu A_\mu$, $\bar \Lambda = \xi^\mu \bar A_\mu$. The infinitesimal conserved charge is
\bea
\sdelta Q_\xi =- \frac{k}{2\pi}\int_0^{2\pi} d\phi\; \xi^\mu \hbox{Tr} \left( A_\mu \delta A_\phi - \bar A_\mu \delta \bar A_\phi \right).
\eea
The boundary conditions require $\xi^\mu$ to depend on $t^+$ only and moreover
\bea
\sdelta Q_\xi &=& \delta \left[ \frac{1}{2\pi}\int_0^{2\pi}d\phi \left( \xi^- \Delta(1+2\p_+\bar P) +\xi^+ \big(\bar L(t^+)-\Delta (\p_+\bar P)^2 \big) \right) \right].
\eea
 which for $\xi^-=\sigma,~\xi^+=\epsilon$ agrees with the integrable charges \eqref{ch1}-\eqref{ch2} in the metric formalism. We have therefore  reduced the charges to the ones discussed in Section \ref{sec:chargesmetric}.

\subsection{From Chern-Simons to constrained WZW}
\subsubsection{The right $\bar A$ sector}
The boundary conditions on $\bar A$ are identical to those of Brown and Henneaux and the analysis of the resulting boundary theory has been given in \cite{Coussaert:1995zp}. Here we quickly review this work in our notation. The variation of the right moving sector is
\bea
\delta S_{-k}[\bar{A}]&=& {k\over2\pi} \int_\Sigma dtd\phi \hbox{Tr}\left(\bar A_t \delta \bar A_\phi \right).\label{csL}
\eea
The conditions (\ref{BC_A2}) imply that on the boundary .\bea
\bar A_- = O(r^{-2}),%\qquad \bar{A}_-\delta \bar{A}_+  = 0
\label{bcAt}
\eea
which enforce $\bar A_t = \bar A_\phi$ close to the boundary. A good variational principle follows from the complete action
\bea
S_R[\bar{A}] &=&  S_{-k}[\bar{A}] -{k\over4\pi}\int_\Sigma d^2x \hbox{Tr}\left({\bar A_\phi^2 - \frac{1}{2}(\bar A_t  - \bar A_\phi)^2}\right).
%&=&S_{CS}[\bar{A}]-{k\over4\pi}\int_\Sigma d^2x \hbox{Tr}\left({\color{Wei}(\bar{A}_+)^2-(\bar{A}_-)^2-2\bar{A}_+\bar{A}_-}\right) .\label{sL}
\eea
We chose to add a combination of the constraints to the action in order to obtain an action of second order in time derivatives, see below \eqref{act}.
The constraint $\bar{F}_{r\phi}=0$ is solved by
\be \bar A_i=G_R^{-1}\p_i G_R,\quad i=r,\phi ,\quad  {G}_R\sim {g}_R(t^+,t^-) \left( \begin{array}{cc}
  {1\over\sqrt{r}}& 0 \\
 0 & \sqrt{r}\end{array}
 \right),\label{defg2}\ee
where the last equation follows from the boundary conditions and $g_R(t^+,t^-)$ is an element of $SL(2,\mathbb R)$. We find
\be{4\pi \over k}S_R[\bar{A}]=\int  \tr {1\over3}({G}_R^{-1}d{G}_R)^3 + {{2}\int_\Sigma dtd\phi \hbox{Tr}\left(\bar{A}_+\bar{A}_- \right)}\ee where $\dot{g}_R\equiv g_R^{-1} \p_t g_R$, $g_R' \equiv g_R^{-1}\p_\phi g_R$\label{acR1}.
It is convenient to use the Gauss decomposition
\bea
G_R(r,t^+,t^-) = \left[ \begin{array}{cc} 1& \hat X_R\\ 0 & 1\end{array}\right] \left[ \begin{array}{cc} \exp(\frac{1}{2}\hat \Phi_R)& 0\\ 0 & \exp{(-\frac{1}{2}\hat \Phi_R)}\end{array}\right]\left[ \begin{array}{cc} 1& 0\\ \hat Y_R & 1\end{array}\right].\label{GaussR}
\eea
One then has
\bea
\frac{1}{3}\tr{(G_R^{-1}dG_R)^3}= d^3x \eps^{\alpha\beta\gamma}\p_\alpha \left[ e^{-\hat \Phi_R} \p_\beta \hat X_R \p_\gamma \hat Y_R \right].  \label{ac1}
\eea
The boundary conditions imply the leading behavior
\bea
\hat X_R(r,t^+,t^-) &\sim& X_R(t^+,t^-),\qquad \hat Y_R(r,t^+,t^-) \sim \frac{1}{r}  Y_R(t^+,t^-),\\
\hat \Phi_R(r,t^+,t^-)&\sim&-\log r + \Phi_R(t^+,t^-).
\eea
The element $g_R(t^+,t^-)$ defined in \eqref{defg2} is then also in the Gauss decomposition with fields $X_R(t^+,t^-)$, $Y_R(t^+,t^-)$, $\Phi_R(t^+,t^-)$.
Hence, the right moving action becomes \bea \label{act}
S_R[\Phi_R,X_R,Y_R]  &=&{k\over8\pi}\int_\Sigma dt^+dt^- ( \p_+\Phi_R\p_-\Phi_R+4e^{-\Phi_R}\p_+ X_R\p_-Y_R ).\nonumber \\
\eea
The tangential components $a = t^+,t^- $ behave as
\bea
\bar A_a \sim \left( \begin{array}{cc} \frac{\bar a_{(3)a}}{2} & \bar a_{(+)a} r \\ -\frac{\bar a_{(-)a}}{r} & -\frac{\bar a_{(3)a}}{2}\end{array}\right) ,\label{au}
\eea
where $\bar{a}_a={g}_R^{-1}\p_a {g}_R$. The boundary conditions further imply
\bea
\bar a_{(3)a} = 0 ,\qquad \bar a_{(\pm)-} = 0, \qquad \bar a_{(+)+} = 1.\label{bcR}\label{eq:18}
\eea
The boundary conditions  (\ref{bcR}) are equivalent to the constraints \bea
0&=& Y_R + \frac{1}{2}\p_+ \Phi_R, \nn\\
0&=&\p_- \Phi_R, \nn\\
1&=&e^{-\Phi_R}\p_+ X_R, \label{consR} \\
0&=& e^{-\Phi_R} \p_- X_R,\nn\\
0&=& \p_- Y_R,\nn
% \\
% \bar{L}&=&{\color{Wei}{k\over4}\left((\p_+\Phi_R)^2-\p_+^2\Phi_R\right)}\nn
\eea
which automatically enforce the equations of motion. We see that all the fields are determined by one right-moving scalar field $\Phi_R(t^+)$:
\bea
\Phi_R = \Phi_R(t^+),\qquad \p_+ X_R(t^+)=e^{\Phi_R(t^+)},\qquad Y_R = - \frac{1}{2}\p_+\Phi_R(t^+).
\eea
The field $\Phi_R$ is subject only to the constraint $\p_-\Phi_R=0$.

The classical Dirac bracket implies that
\be \{\p_+\Phi_R(t^+),\p_+\Phi_R(t'^+)\} =-  {\frac{4\pi}{k}}\p_{t^+}\delta(t^+-t'^+)
\ee
%{\color{Wei} Note that the minus sign comes from the overall minus in front of $S_R$ in the total action (\ref{SE}).}
The stress tensor following from the action (\ref{act}) is
\be T^{R-}\,_+={k\over 4}( \p_+\Phi_R\p_+\Phi_R+4e^{-\Phi_R}\p_+X_R\p_+Y_R ).\ee
Using (\ref{consR}) this becomes the stress-tensor of a linear dilaton CFT,
\be\label{cr} T^{R-}\,_+={k\over 4}( \p_+\Phi_R\p_+\Phi_R-2\p^2_+\Phi_R ).\ee
We can then read off the classical central charge
\be c_R=6k.\ee
We also note that $T^{R\pm}\,_-=T^{R+}\,_+=0$.We finally note from (\ref{BC_A2}) that in terms of metric components we have the on-shell relation
\be T^{R-}\,_+= \bar L(t^+).\ee
\subsubsection{The left $A$ sector}
 The boundary conditions in the unbarred  sector are different from those of Brown and Henneaux. In the
 usual case we would take $\delta A_+=0$ and the unbarred sector completes the barred sector to a
 nonchiral Liouville theory. However this condition is incompatible with the new boundary condition
 (\ref{BC_A1}). The variation of the action reads as
\bea
\delta S_{k}[A]= \frac{1}{2\pi}\int_\Sigma dtd\phi\left( \delta \Delta-\delta(\Delta \p_+\bar P^2(t^+))
+2\Delta\delta \p_+\bar P(t^+) \right).\label{variation}
\eea
A good variational principle arises from the action\footnote{One can also add a boundary term proportional to $\int dt^+dt^- \tr (A_-^2)$ since its variation is proportional to $\delta \Delta =0$. The only effect of such term is to shift the current $T^{L-}_{\;\;\; -}$ defined below by a constant.}
\bea
S_L[A] =S_{k}[A] - \frac{k}{8\pi}\int_{\Sigma}dt^+dt^-  \tr (A_-^2-A_+^2-2A_+A_--4\alpha
[A_+,A_-]\sigma^{(3)}).\nn\\
\eea
where $\alpha $ is a constant since
\bea
\delta S_L = -\frac{1}{2\pi}\int dt^+dt^- \p_+\bar P \delta\Delta = 0\, .
\eea 
With Brown-Henneaux boundary condition we should take $\alpha=0$. With the
new boundary conditions, we will see later that we should choose $\alpha=1$.
The constraint $F_{r\phi}=0$ can be solved consistently with the boundary conditions by
\bea
A  = G_L^{-1}\p_i G_L,\qquad i=r,\phi,\qquad
G_L \sim g_L(t^+,t^-) \left( \begin{array}{cc} \sqrt{r}&0 \\ 0&
\frac{1}{\sqrt{r}}\end{array}\right),\label{as2}
\eea
where $g_L(t^+,t^-)$ is an element of $SL(2,\mathbb R)$. The action can be written as
\bea
S_{L}[A]&=&{k\over4\pi}\left(-\int d^3x{1\over3}(G_L^{-1}dG_L)^3 + \half \int_\Sigma dt d\phi
\hbox{Tr}\left(\dot{g}_L^2- (g_L')^2\right)\right),\nn
\eea
where $\dot g_L \equiv a_t$, $g_L'\equiv a_\phi$ with $a_a = g_L^{-1}\p_a g_L$. Using the Gauss
decomposition
\be g_L= \left( \begin{array}{cc}
 1 & 0\\
Y_L & 1\end{array}
 \right) \left( \begin{array}{cc}
e^{-\Phi_L\over2} &0 \\
0&e^{\Phi_L\over2}\end{array}
 \right)
\left( \begin{array}{cc}
 1 & X_L\\
0 & 1\end{array}
 \right),\label{GaussL}
\ee
the action becomes
\bea
S_{L}[X_L,Y_L,\Phi_L]
&=& {k\over8\pi}\int_\Sigma dt^+dt^-  [\p_+\Phi_L\p_-\Phi_L+4e^{-\Phi_L}\p_+ X_L\p_-Y_L\label{actL} \\
&&\hspace{-2cm}+4\alpha
e^{-\Phi_L}\left(X_L(\p_+Y_L\p_-\Phi_L-\p_+\Phi_L\p_-Y_L)+\p_+X_L\p_-Y_L-\p_-X_L\p_+Y_L\right)] . \nn
\eea
For Brown-Henneaux boundary conditions, the constraints are
\bea
\p_+ X_L=\p_+ Y_L = \p_+ \Phi_L = 0,\\
X_L = -\frac{1}{2}\p_- \Phi_L,\qquad \p_- Y_L = e^{\Phi_L}.
\eea
The left-moving chiral boson is then expressed as $\frac{k}{4}((\p_- \Phi_L)^2 - 2 \p_-^2 \Phi) =  L(t^-)$ and the left action with $\alpha = 0$ is similar to the right action \eqref{act} with right fields replaced by left fields. Note the Gauss decomposition \eqref{GaussR}-\eqref{GaussL} was instrumental in describing the left and right sector with similar variables.

Under the new boundary conditions, using
\bea
A_a \sim \left[ \begin{array}{cc} \frac{a_{(3)a}}{2} &\frac{a_{(+)a}}{r}  \\ -a_{(-)a} r&\quad
-\frac{a_{(3)a}}{2}\end{array}\right] , \qquad a=+,-
\eea
we see that (\ref{BC_A1}) implies
\bea
a_{(3)a} = 0,\qquad  a_{(-)-}=-1, \qquad a_{(-)+} = \p_+\bar P,\nn\\
a_{(+)-} = \frac{\Delta}{k},\qquad a_{(+)+} = -\frac{\Delta}{k}a_{(-)+}.\label{eq:15}
\eea
In terms of the Gauss decomposition this becomes
\bea
\p_+ X_L&=&( X_L^2-{\Delta\over k})\p_+\bar P,\\
\p_- X_L&=&{\Delta\over k}-X_L^2,\\
\p_- \Phi_L&=&-2X_L,\\
\p_+ \Phi_L &=&2\p_+\bar PX_L,\\
\p_+ Y_L &=&-e^{\Phi_L}\p_+\bar P, \\
\p_- Y_L&=&e^{\Phi_L} .
\eea
These equations determine $\Phi_L$, $X_L$ and $Y_L$ in terms of $\Delta$ and $\p_+\bar P(t^+)$. The solutions to the constraints is
\bea
X_L&=&\sqrt{\Delta\over k}\tanh\left( \sqrt{\Delta\over k}(t^- -\bar P)\right),\nn\\
e^{\Phi_L}&=&e^{\phi_0}\cosh^{-2}\left(\sqrt {\Delta\over k}(t^- -\bar P)\right),\\
Y_L&=&y_0+e^{\phi_0}\sqrt{k\over\Delta}\tanh\left(\sqrt {\Delta\over k}(t^-  -\bar P)\right).\nn
\eea
This solution is periodic in $\phi = (t^+-t^-)/2$ when $\Delta = -k/4$ (which corresponds to the AdS vacuum). It is periodic in $\phi$ for $\Delta >0$ only when $\bar P = -t^+ + f(t^+)$ where $f$ is a periodic function.

Using the constraint conditions and choosing $\alpha=1$, one can write the action as
\bea \hspace{-18pt} S_{L}[X_L,Y_L,\Phi_L]
&=& {k\over 8\pi}\int_\Sigma dt^+dt^-  [ -(\p_+\Phi_L\p_-\Phi_L+4e^{-\Phi_L}\p_+X_L\p_-Y_L)\nn\\&&+8{\Delta\over
k}{\p_+X_L\over {\Delta\over k}- X_L^2} ].\label{actLf}
\eea
One can check that the equations of motion from (\ref{actLf}) are fully compatible with the constraints. (The second line in the action does not contribute to the equations of motion.) Therefore, it is correct to use the action (\ref{actLf}). The action of the right moving conformal transformations on these fields is nontrivial because they depend
on $t^+$.
Under the reparameterization $\delta t^+= \xi(t^+)$,  $\Phi_R,\,Y_R,X_R$ transform as scalars, and
$\p_+\bar P$ transform as a weight one operator \be\delta_\xi(2 \Delta \p_+\bar P(t^+))=-\p_+\xi
(2\Delta \p_+\bar P(t^+))-\xi \p_+(2\Delta \p_+\bar P(t^+)).\ee

One can construct the stress-tensor as the Noether current
\bea  T^{L-}\,_+(t^+)&=&-{k\over4}\left((\p_+\Phi_L)^2+4e^{-\Phi_L}\p_+X_L\p_+Y_L\right)\nn\\
&=& {1\over k_{KM}}(2 \Delta \p_+\bar P )^2,
\eea
together with $T^{L+}\,_{+}(t^+)=0$. Summing up the contributions from the left and right sectors, we
obtain that there is no correction to the central charge obtained from the right sector $c=6k$ (although
at the quantum level there would be a shift from $c=6k+1$ to $c=6k+2$).

Now we wish to construct the Noether current associated to the
crossover symmetry (\ref{crs}) under which  $\delta_\sigma t^-=\sigma(t^+)$. The fields transform as
\bea
\delta_\sigma \Phi_L =-\sigma\p_-\Phi_L,\qquad
\delta_\sigma Y_L =-\sigma\p_-Y_L,\qquad
\delta_\sigma X_L =-\sigma\p_-X_L,
\eea
or equivalently
\be\delta_\sigma (2\Delta \p_+\bar P)=2\Delta\p_+\sigma .
\ee
This is generated by
\bea T^{L-}\,_{-}(\sigma)&=&-{k\over2}\left((\p_+\Phi_L\p_-\Phi_L+4e^{-\Phi_L}\p_+X_L\p_-Y_L\right)\nn\\
&=&2\Delta \p_+\bar P(t^+),  \eea
whose commutators form a Kac-Moody algebra at level $ k_{KM}$ as anticipated.
The zero mode of the left translation is generated by
\bea T^{L+}\,_{-}(\sigma)&=&-{k\over4}\left((\p_-\Phi_L\p_-\Phi_L+4e^{-\Phi_L}\p_-X_L\p_-Y_L-8{\Delta\over
k}{\p_-X_L\over{\Delta\over k}- X_L^2} \right)\nn\\ &=&\Delta.
\eea
The  combined charge
\bea \mathcal{Q}_\xi&\equiv&  {1\over2\pi}\int_0^{2\pi} d\phi\Big( (T^{R+}\,_++T^{L+}\,_+)\epsilon+(T^{L+}\,_-+T^{L-}\,_-)\sigma\Big)%&=& -\int d\phi T_{R++}(e^{int^+})+\left(T_{L++}- T_{L+-} + T_{L-+} - T_{L--}\right)(e^{int^+})\nn
\eea % where $n_\pm = \half$ ,
generates the diffeomorphism $\xi^+=\epsilon,\,\xi^-=\sigma$,  agreeing with the charges computed in the gravity calculation (\ref{ch1}) and (\ref{ch2}).

\subsection{From chiral bosons to Liouville }
\subsubsection{Brown-Henneaux boundary conditions}
For the case of Brown-Henneaux boundary conditions,
our analysis determined all the on shell fields in terms of two free bosons $Y_L(t^-)$ and $X_R(t^+)$. The other fields are given by
\bea
e^{-\Phi_L}\p_-Y_L&=&1,\quad X_L=-\half\p_-\Phi_L,\\
e^{-\Phi_R}\p_+X_R&=&1,\quad Y_R=-\half\p_+\Phi_R,
\label{sl}\eea
where $(X,Y,\Phi)_{L/R}$ parameterize the group element $g_{L/R}$ by
\bea
g_L&=&\left[ \begin{array}{cc} 1& 0\\ Y_L & 1\end{array}\right] \left[ \begin{array}{cc} \exp(-\frac{1}{2} \Phi_L)& 0\\ 0 & \exp{(\frac{1}{2} \Phi_L)}\end{array}\right]\left[ \begin{array}{cc} 1& X_L\\ 0 & 1\end{array}\right]\label{gaussL},\\
g_R&=&\left[ \begin{array}{cc} 1& X_R\\ 0 & 1\end{array}\right] \left[ \begin{array}{cc} \exp(\frac{1}{2} \Phi_R)& 0\\ 0 & \exp{(-\frac{1}{2} \Phi_R)}\end{array}\right]\left[ \begin{array}{cc} 1& 0\\ Y_R & 1\end{array}\right].
\eea
The Liouville field $\Phi$ comes from the Gauss decomposition of $g\equiv g_L^{-1}g_R$,
\be g=\left[ \begin{array}{cc} 1& X\\ 0 & 1\end{array}\right] \left[ \begin{array}{cc} \exp(\frac{1}{2} \Phi)& 0\\ 0 & \exp{(-\frac{1}{2} \Phi)}\end{array}\right]\left[ \begin{array}{cc} 1& 0\\ Y & 1\end{array}\right].
\ee
Therefore
\bea e^{-\half\Phi}=e^{-\half (\Phi_L+\Phi_R)}(1-X_RY_L)\label{phi} .\eea
Plugging the solution (\ref{sl}) into the above definition, we get the relation
\be e^{\Phi}={\p_+X_R(t^+)\p_-Y_L(t^-)\over (1-X_R(t^+)Y_L(t^-))^2},\ee
known as a B\"acklund transformation. 
$\Phi$ satisfies the Liouville field equation
\be
\p_+\p_-\Phi-2e^{\Phi}=0.
\ee
Hence the two free fields are equivalent to the Liouville scalar.

\subsubsection{New boundary conditions}

Using  the field $\Phi$ defined in  (\ref{phi}), we still find that
\be  e^\Phi={\p_-Y_L\p_+X_R\over \left ( 1-Y_LX_R\right)^2} \ee
but $\Phi$ and $h\equiv \p_+\bar P$ are solutions of the following constrained system
\bea
S&=&{k\over 8\pi}\int dt^+dt^- \left(\p_+\Phi\p_-\Phi+4e^\Phi+h( (\p_-\Phi)^2-2\p_-^2\Phi-{4\Delta\over k})\right),\nn\\
\label{aclv}\\
\p_-h&=&0.
\eea
The right-moving current $\bar{T}^-\,_+$ that generates  the  diffeomorphism $t^+\rightarrow t^++\epsilon(t^+)$ is
\bea \bar{T}^-\,_+&=&{k\over4}\left((\p_+\Phi)^2-2\p_+^2\Phi+2h(\p_+\Phi\p_-\Phi-2\p_+\p_-\Phi)-2\p_+h\p_-\Phi\right),\nn\\
\p_-\bar T^-\,_+&=&0.
\eea
The  right-moving current that generates the diffeomorphism $t^-\rightarrow t^-+\sigma(t^+)$ is
\bea
\bar{T}^-\,_-&=&{k\over2}h\left((\p_-\Phi)^2-2\p_-^2\Phi\right)=2\Delta h\\
\bar{T}^+\,_-&=&{k\over4}\left((\p_-\Phi)^2-2\p_-^2\Phi\right)=\Delta
\eea
The on shell results agree with the previous subsection, and reduce to the standard Liouville stress tensors at $h=0.$
The currents generate the symmetry transformations via the Dirac brackets \bea
\{2\Delta h(t^+) ,\, 2\Delta h({s^+})\}&=&{\pi k_{KM}}\p_{t^+}\delta(t^+-{s^+})\label{hh},\\
\{2\Delta h(t^+) ,\, \Phi(s^+,s^-)\}&=&-2\pi\delta(t^+-s^+)\p_{s^-}\Phi(s^+,s^-),\\
\{\bar{T}^-\,_+(t^+),\,h({s^+})\}&=&-2\pi\p_{s^+}\left(\delta(t^+-s^+)h(s^+,s^-)\right),\\
%\{2\Delta \beta(t^+),\,\Phi(s^+,s^-)\}&=&-2\pi\delta(t^+-s^+)\p_{s^-}\Phi(s^+,s^-)\\
\{\bar{T}^-\,_+(t^+),\,\Phi(s^+,s^-)\}&=&-2\pi\left(\delta(t^+-s^+)\p_{s^+}\Phi(s^+,s^-)-\p_{t^+}\delta(t^+-s^+)\right),\nn
\\
%\{\Phi(t^+,t^-),\,\Phi(s^+,s^-)\}&=&
\{\bar{T}^-\,_+(t^+),\,\bar{T}^-\,_+({s^+})\}&=&2\pi(\p_{t^+}-\p_{s^+})\left(\delta(t^+-s^+)\bar T(s^+,s^-)\right)\nn\\
&&-{\pi c_R\over 6}\p_{t^+}^3\delta(t^+-s^+).\label{tt}
\eea
The first line  is just the $U(1)$ Kac-Moody algebra with level $\tilde{k}_{KM}=-4\Delta$. The last is the  Virasoro algebra with central charge $c_R$. 

In the companion paper \cite{css2}  we derive the action (\ref{aclv}) (in the notation $\Phi=2\rho$) by imposing a chiral gauge condition on  the Polyakov-Liouville  action for two-dimensional gravity. The equations of motion agree and applying the canonical formalism to this  chiral Liouville theory gives the Dirac brackets (\ref{hh})-(\ref{tt}), with the identification $\bar{T}^-\,_+=j^-_\epsilon,\, \bar{T}^-\,_-=j^-_\sigma\,,$ and $c=c_R=6k$. Therefore the two-dimensional chiral Liouville theory and AdS$_3$  Chern-Simons gravity with the new boundary conditions are semiclassically equivalent.

\section{%Lowest-weight
Bulk Kac-Moody representations }

In this section we describe the lowest-weight Kac-Moody representations in terms of a weakly interacting gas of bulk particles.
We work in the global AdS$_3$ coordinates\be
{ds^2 \over \ell^2} = - \cosh^2 \rho d\tau^2 + \sinh^2 \rho d\phi^2 + d \rho^2.
\ee
In these coordinates, with $u = \tau + \phi$, $v = \tau-\phi$,
the $\overline{SL(2,{\mathcal R})}_R$ generators are described by the vector fields
\bea\label{vr}
\bar \cL_0 &= &  i \partial_u ~, \cr
\bar \cL_{-1} &= &  i e^{-iu} \left[ { \cosh 2 \rho \over \sinh 2 \rho } \partial_u
- { 1 \over \sinh 2 \rho} \partial_v +  { i \over 2} \partial_\rho
\right] ~, \cr
\bar \cL_{1} &= &  i e^{iu} \left[ { \cosh 2 \rho \over \sinh 2 \rho } \partial_u
- { 1 \over \sinh 2 \rho} \partial_v -  { i \over 2} \partial_\rho
\right]~, \eea
normalized so that
\be
[\bar \cL_0,\bar \cL_{\pm1} ] = \mp  \bar \cL_{\pm 1}~,~~~~~~~~~~~~[\bar \cL_1,\bar \cL_{-1} ] = 2 \bar \cL_0~.
\ee
The quadratic Casimir of $SL(2,{\mathcal R})_R$ on scalar fields is
\be
\bar \cL^2 = { 1\over 2} ( \bar \cL_{1} \bar \cL_{-1} + \bar \cL_{-1} \bar \cL_1 ) - \bar \cL_0^2 =-{1 \over 4} [ \partial_\rho^2 + 2 { \cosh 2\rho \over
\sinh 2 \rho } \partial_\rho  + {1 \over \sinh^2 \rho} \partial^2_\phi
- { 1 \over \cosh^2 \rho } \partial_\tau^2 ]~,
\ee
which is the laplacian times $-\ell^2/4$. Therefore a scalar field of mass
$m$ has
\be{\bar \cL^2 =-m^2 \ell^2/4 ~,}\ee
and the conformal algebra can be used to classify the solutions of
the wave equation. We also have a $U(1)$ isometry
\be\bar \cJ_0=i\p_v,\ee sometimes called $\cL_0$.
In the usual manner consider states (i.e. scalar field wavefunctions) with weights $(h,p)$ under
$\bar \cL_0,\bar \cJ_0$ so that
\be
\bar \cL_0 | \psi \rangle = h  | \psi \rangle ~,~~~~~~~~~~
\bar \cJ_0 | \psi \rangle = p  | \psi \rangle ~.
\ee
It follows that
\be
| \psi \rangle = e^{- i h u  -i p v } F(\rho) ~.
\ee
Periodicity requires
\be h-p\in {\bf Z} .\ee
Now suppose that $| \psi \rangle$
is a $SL(2,{\mathcal R})_R$ primary state in the
sense that $\bar \cL_1 | \psi \rangle =  0$.
This implies that
 $F$ satisfies
\be
2h { \cosh 2\rho \over \sinh 2\rho }F -{2p \over \sinh 2 \rho}F+  \partial_\rho F  =0 ~,
\ee
which is solved by
\be \label{fd}
F(p,h) = const { (\sinh\rho)^{p-h} \over ( \cosh \rho )^{p+h} } ~.
\ee
Regularity at the origin requires \be p\ge h. \ee
Demanding that $ | \psi \rangle$ obey the mass $m$ wave equation
and be normalizable at large $\rho$   with Dirichlet boundary conditions imposes the additional constraint
\be h = {1 \over 2}\left(1  +
 \sqrt{{m^2 \ell^2 }+ 1} \right) ~.\ee

 Starting from these primary states labelled by $p\ge h$ we can generate all other
normalizable solutions with Dirichlet boundary conditions at
infinity by acting with $\bar \cL_{-1}$.
These descendants all have the same quadratic Casimirs,
but higher   integer-spaced eigenvalues of $\bar \cL_0$.
They correspond in an obvious way to $\overline{SL(2,{\mathcal R})}_R$ descendants
of the primary operators.

Recall the usual procedure with Brown-Henneaux boundary conditions in AdS$_3$ is to also use the $SL(2,{\mathcal R})_L$ isometries (which look like (\ref{vr}) with a $u\leftrightarrow v$ exchange). All solutions then arise from left-right primaries and their descendants. The states above correspond to acting on the primary with $p=h$ with $\cL_{-1}^{p-h}$:
\be F(p,h)= \cL_{-1}^{p-h}F(h,h).\ee  With the usual Brown-Henneaux boundary conditions, the left-right primary wavefunction is (\ref{fd})
with $p=\bar h=h$. The full set of Virasoro descendents are then all states of the form
\be\label{xz} \left((\bar \cL_{-{\bar k}})^{n_{\bar k}}....(\bar \cL_{-2})^{n_{\bar 2}}(\bar \cL_{-1})^{n_{\bar 1}}\right) \left((\cL_{-k})^{n_k}....(\cL_{-2})^{n_2}(\cL_{-1})^{n_1}\right)F(h,h)\ee
This state has
\be \cL_0=h+\sum_{j=1}^k jn_j,~~~~\bar \cL_0=h+\sum_{\bar j=1}^{\bar k} {\bar j}n_{\bar j}.\ee
Using the comutation relations, it is not hard to see that we can always arrange to move the higher moded operators to the left.
%: i.e. $m_{f}\ge m_{f-1}$ etc.
The particle interpretation of this is as follows. The original primary is a minimal energy single scalar particle sitting in the middle of AdS$_3$. The action of powers of $\bar \cL_{-1}$ and $\cL_{-1}$ serve to translate its position around to all possible locations in  AdS$_3$.  $\bar \cL_{-2}$ and $\cL_{-2}$ create left and right boundary gravitons in their minimal energy states, and the higher Virasoro generators create excited boundary gravitons. Hence the left-right Virasoro module corresponds to multi-particle states consisting of a single scalar particle accompanied by a gas of boundary gravitons. The standard character is
\be tr[q^{\cL_0}\bar q^{\bar \cL_0}]={q^{h-\frac{1}{24}}\bar q^{h-\frac{1}{24}} \over \eta(q) \eta(\bar q)}.\ee

However with the new chiral boundary conditions $\cL_{-n}$ is not part of the asymptotic symmetry group (except for $\cL_0=\bar \cJ_0$) so these solutions do not all obey the new boundary  conditions. Using the commutation relations the general state obeying the new boundary conditions can be written in the form
\be \left((\bar \cJ_{-{\bar l}})^{q_{\bar l}}....(\bar \cJ_{-2})^{q_{\bar 2}}(\bar \cJ_{-1})^{q_{\bar 1}}\right) \left((\bar \cL_{-k})^{n_k}....(\bar \cL_{-2})^{n_2}(\bar \cL_{-1})^{n_1}\right)F(h,p)\ee
This has
\be \bar \cL_0=h+\sum_{j=1}^k jn_j+\sum_{\bar j=1}^{\bar l} \bar j q_{\bar j},~~~\bar \cJ_0=p. \ee
We see that instead of right and left moving gravitons we have  right moving gravitons and photons. The number of states is roughly the same: for every state in the non-chiral theory we can get one in the chiral theory by replacing
$p=h+n_{\bar 1}$ and $q_{\bar l}=n_{\bar l+1}$. The chiral character is
 \be tr[\bar q^{\bar \cL_0}\bar \mu^{\bar \cJ_0}]={\bar q^{h-\frac{1}{12}} \bar \mu^h  \over \eta^2(\bar q) (1-\bar \mu)}.\ee

\section{String theory and warped $AdS_3$}
\label{sts}

String theory is an excellent source of fully consistent theories of Einstein gravity with matter on AdS$_3$. Usually these are taken to obey standard Brown-Henneaux boundary conditions. In this section we consider the string solutions studied in \cite{Azeyanagi:2012zd}
for  which the new boundary conditions may be relevant. These solutions generically contain a warped AdS$_3$ factor, but reduce to AdS$_3$ in an appropriate limit. We will see that a natural set of boundary conditions for the generic case reduce to our new boundary conditions in the AdS$_3$ limit.

The universal massless NS-NS sector of string theory compactified to six dimensions is governed by the low energy effective lagrangian %in Einstein frame ($g_E = e^{-\phi}g_S$)
\bea
L = \frac{1}{16\pi G_6}\sqrt{-g}e^{-2\phi}(R+4(\p\phi)^2-\frac{1}{12}H^2).\label{NSL}
\eea The six dimensional
warped AdS$_3 \times$S$^3$ black string solution discussed in \cite{Azeyanagi:2012zd} is
\bea
{4ds_{6}^2\over \ell^2}&=&-{\rho^2-(2\pi^2T_L T_R)^2\over \pi^2 T_L^2}(dt^+)^2+{d\rho^2 \over \rho^2-(2\pi^2T_L T_R)^2}+4d\Omega_3^2\nn\\
&& \hspace{-0.6cm}+ 4\pi^2 T_L^2\Xi^2+8\pi T_L{{\sinh\alpha\over\cosh^2\alpha}}\Xi(d\chi+\cos\theta d\phi),\nn\\
B&=&-{\ell^2\over{4}} \Big( \cos\theta d\phi \wedge d\chi+2\rho \Xi \wedge dt^++4\pi T_L{{\sinh\alpha\over\cosh^2\alpha}} (d\chi+\cos\theta d\phi)\wedge\Xi \Big)\label{TsTBTZ},\nn\\
e^{-2\phi}&=&\cosh^2\alpha,
\eea
where \bea
 \sinh\alpha &=&\lambda \pi T_L, \qquad \Xi = d\tau^- -{\rho dt^+\over 2\pi^2 T_L^2},\\
 4d\Omega_3^2&=&(d\chi+\cos\theta d\phi)^2+d\theta^2+\sin^2\theta d\phi^2.
 \eea
We take $(\pi \lambda T_L)^2 < 1$. The coordinate $\psi = \chi + 2\pi T_L \sinh\alpha\, \tau^-$ is identified with period $4\pi$ while $t^- = \tau^- + \frac{\lambda}{2} \chi$ is unidentified. Equivalently $\chi\sim \chi +{4\pi\over 1-\sinh^2\alpha} $.
%Using the methods of \cite{Barnich:2001jy,Barnich:2007bf,Compere:2007az}, the conserved charge densities per unit $dt/l$ are given by
%\bea
%\mathcal Q_{\p_{t_+}} = \frac{\pi^2 l }{6} c_R T_R^2, \qquad \mathcal Q_{\p_{t_-}} = \frac{\pi^2 l }{6} c_R \frac{T_L^2}{1-\pi^2 \lambda^2 T_L^2}.
%\eea
%where $c_R=3l/2G_3$ and $G_6 = 16\pi^2 l^3 G_3$. The divergence at
%\bea
%\pi \lambda T_L = \pm 1
%\eea
%originates from the fact that the metric is then non-invertible. The TsT transformation used to generate the solution is not defined at that values of the parameters.
%
Using the coordinates $(t^+,\tau^-,\rho,\theta,\phi,\chi)$, we impose  the following boundary conditions on the string metric and B field,
\bea
\frac{g_{\mu\nu}+B_{\mu\nu}}{\ell^2}&=&\left(
                \begin{array}{cccccc}
                 \rho \p_+\bar P(t^+) & - \rho &0 & 0 & -\frac{\tanh{\alpha}}{\pi T_L \cosh\alpha}\cos\theta \rho& -\frac{\tanh{\alpha}}{\pi T_L \cosh\alpha} \rho\\
              0   & \pi^2 T_L^2& 0 & 0 & \frac{2\pi T_L \tanh{\alpha}}{ \cosh\alpha}\cos\theta & \frac{2\pi T_L \tanh{\alpha}}{ \cosh\alpha} \\
                  0 &0  & \frac{1}{4\rho^2} & 0 & 0 & 0 \\
                0   &  0& 0 & \frac{1}{4} & 0 & 0 \\
	       0& 0 &  0& 0 & \frac{1}{4} &0 \\
              0   &  0 &  0& 0 & \frac{\cos\theta}{2}  & \frac{1}{4} \\
                \end{array}
              \right)\nonumber\\
              +&&\hspace{-1cm}\left(
                \begin{array}{cccccc}
                \mathcal{O}(\rho^0)     &\mathcal{O}(\rho^{-1})                    &\mathcal{O}(\rho^{-2})  & \mathcal{O}(\rho^{0}) & \mathcal{O}(\rho^{0}) & \mathcal{O}(\rho^{0}) \\
                \mathcal{O}(\rho^{0})   &\mathcal{O}(\rho^{-1})                  & \mathcal{O}(\rho^{-2}) & \mathcal{O}(\rho^{-1}) & \mathcal{O}(\rho^{-1}) & \mathcal{O}(\rho^{-1}) \\
                \mathcal{O}(\rho^{-2})  &\mathcal{O}(\rho^{-2})                  & \mathcal{O}(\rho^{-3}) & \mathcal{O}(\rho^{-2}) & \mathcal{O}(\rho^{-2}) & \mathcal{O}(\rho^{-2}) \\
                \mathcal{O}(\rho^{0})   &\mathcal{O}(\rho^{-1})                &  \mathcal{O}(\rho^{-2}) & \mathcal{O}(\rho^{-1}) & \mathcal{O}(\rho^{-1}) & \mathcal{O}(\rho^{-1}) \\
                \mathcal{O}(\rho^{0})   &\mathcal{O}(\rho^{-1})                    &  \mathcal{O}(\rho^{-2})  &  \mathcal{O}(\rho^{-1})  & \mathcal{O}(\rho^{-1}) & \mathcal{O}(\rho^{ -1}) \\
                \mathcal{O}(\rho^{0})   &\mathcal{O}(\rho^{-1})                   & \mathcal{O}(\rho^{-2}) &   \mathcal{O}(\rho^{-1})  &   \mathcal{O}(\rho^{-1}) & \mathcal{O}(\rho^{-1}) \\\end{array}
              \right)\label{bcwarped}
\eea
where all components are also allowed to depend upon $(\theta,\phi,\psi)$. Note that we impose $(g+B)_{+-} =O(\rho^{-1})$ while $(g+B)_{-+}  = O(\rho^0)$. The asymptotic symmetry group is generated by the following combination of asymptotic vectors and large gauge transformations,
\bea
(\xi_m,\Lambda^\xi_m)&=&e^{i m t^+}\Big(\p_+ -{\p_-}- im \rho \p_\rho-\frac{m^2}{2\rho}\p_-+O(\rho^{-2})\p_\pm + O(\rho^{-1})\p_r,\nn\\
&&  l^2 \pi^2 T_L^2 d\tau^- - \frac{il^2 m}{4\rho}d\rho\Big) ,\nn\\
(\eta_m,\Lambda^\eta_m)&=&e^{imt^+} \Big( {\sqrt{2G_6}\over\pi^2 T_L\ell^2}\p_-,  \sqrt{2G_6}T_Ld\tau^-\Big) \label{asKstring},\\
(\eta^a_m,\Lambda^\eta_m )&=&  e^{imt^+}\Big(k^a, \Lambda^{k^a}\Big), \nonumber\\
(\bar{\eta}_m,\Lambda^{\bar\eta}_m) &=&e^{imt^+} \Big( \p_\chi , \frac{l^2}{4}d\chi \Big),\nn
\eea
where  $\p_+ = \p_{t^+}$, $\p_- = \p_{\tau^-}$ and  $k^a$ are $SU(2)_L$ generators
\be k^3=-i\p_\phi,\quad k^\pm=e^{\mp i\phi}(\p_\theta\mp i(\cot\theta\p_\phi-\csc\theta \p_\chi)),\ee
and
\bea
\Lambda^{k^3} &=& i\frac{l^2}{4}d\phi,\nn\\
\Lambda^{k^\pm}&=&e^{\mp i \phi}\frac{l^2}{4}\Big(-d\theta \mp i ( \cot\theta d\phi +\csc\theta d\chi) \Big).
\eea
It may be shown that the boundary conditions are preserved by the action of the asymptotic symmetry group. The subleading term $\sim \frac{1}{\rho}\p_-$ in $\xi_m$ ensures that $\delta g_{\rho-} = O(\rho^{-2})$.

A subtlety is that the asymptotic generators are not uniquely fixed by the boundary conditions \eqref{bcwarped}. They are constrained with the supplementary conditions
\bea
\delta_{(\xi,\Lambda)} (g+B)_{\phi+} = O(\rho^{-1}), \qquad \delta_{(\xi,\Lambda)} (g+B)_{\chi +} = O(\rho^{-1}) ,\label{mixedgB1}
\eea
for $(\xi,\Lambda) = (\xi_m,\Lambda_m^\xi),\, (\eta_m,\Lambda^\eta_m),\, (\bar{\eta}_m,\Lambda^{\bar\eta}_m),\,  $ while
\bea
\delta_{(\eta^a_m,\Lambda^\eta_m )} (g+B)_{+\phi} = O(\rho^{-1}), \qquad \delta_{(\eta^a_m,\Lambda^\eta_m )} (g+B)_{+\chi} = O(\rho^{-1}) .\label{mixedgB2}
\eea
These mixed boundary conditions can be motivated by the string construction of vertex operators  \cite{Azeyanagi:2012zd}.

The asymptotic charges and asymptotic symmetry algebra can be obtained using the formalism of \cite{Barnich:2001jy,Barnich:2007bf,Compere:2007vx,Compere:2009dp}. In particular, using the $4$ form $k_{\xi,\Lambda} ^{\mu\nu}\eps_{\mu\nu \alpha_1\dots \alpha_4}$ derived from the Lagrangian \eqref{NSL}, the conserved charges are given by \bea \sdelta Q_{\xi,\Lambda} = \int d\sigma d\chi d\phi d\theta (k^{rt^+}_{\xi,\Lambda} +k_{\xi,\Lambda}^{r\tau^-}+\frac{\lambda}{2}k_{\xi,\Lambda}^{r\chi}) \eea where $\sigma = (t^+-t^-)/2$ is taken to be of period $2\pi$. We then obtain the right moving Virasoro central charge
\be c_{R,\lambda}=\frac{3\pi^2 \ell^4}{G_6}\ee
a U(1) Kac-Moody level \be k_{U(1)}=-4\label{KAclev}\ee
an SU(2) Kac-Moody level \be k_{SU(2)}=\frac{\pi^2 \ell^4}{2G_6} \label{SU2k}\ee
and $\overline{U(1)}$ Kac-Moody level \be \bar{k}_{U(1)}=\frac{\pi^2 \ell^4}{2G_6}.\label{U1k}\ee
Accounting for normalization conventions ($2G_6=\pi^2$) this agrees with the results of the world sheet computations  of the space-time central charges given in \cite{Azeyanagi:2012zd}. We note that  \cite{Azeyanagi:2012zd} also gives a construction of the world sheet vertex operators for the boundary photons and gravitons.   In three-dimensional language (after reduction on the three-sphere), the level \eqref{KAclev} is negative for spacelike warped $AdS_3$ black holes ($T_L^2 >0$) and becomes positive for timelike warped spacetimes ($T_L^2 < 0 $) after setting a real normalization in \eqref{asKstring} (because $\eta \rightarrow i \eta$ leads to $k_{U(1)} \rightarrow -k_{U(1)}$). These qualitative features of the $U(1)$ Kac-Moody level are similar to the ones derived earlier in another class of warped geometries \cite{Compere:2008cv,Compere:2009zj}.

The boundary conditions \eqref{mixedgB1}-\eqref{mixedgB2} are crucial in order to derive the large gauge transformations in \eqref{asKstring}. These large gauge transformations then lead to the correct values of the Kac-Moody levels \eqref{SU2k}-\eqref{U1k}. This is an example where matter fields directly contribute to central extensions in the asymptotic symmetry algebra. 
We expect that the boundary conditions \eqref{bcwarped} (with possible mild modifications) will lead to finite, integrable and conserved charges associated with the generators \eqref{asKstring}. This however remains to be fully checked. 

To compare with the preceding sections, we should take the warping factor $\lambda=0$ and look at the 3d Einstein frame metric $g_{\mu\nu}$ defined by
\be ds_{6(s)}^2=e^{4\phi}ds_{3(E)}^2+\ell^2d\Omega_3^2\ee
The boundary conditions (\ref{bcwarped}) reduce to
\bea
\frac{g_{\mu\nu}}{\ell^2}&\sim&\left(
                \begin{array}{cccccc}
                 \rho \p_+\bar P(t^+) & -\frac{\rho}{2} &0  \\
                & {\pi^2 T_L^2} & 0  \\
                   &  & \frac{1}{4\rho^2}  \\
                \end{array}
              \right)+\left(
                \begin{array}{cccccc}
                \mathcal{O}(\rho^0) & \mathcal{O}(\rho^{0}) &\mathcal{O}(\rho^{-2}) \\
                &\mathcal{O}(\rho^{-1}) & \mathcal{O}(\rho^{-2}) \\
                   &  & \mathcal{O}(\rho^{-3})  \\\end{array}
              \right).\label{bcback}\nonumber
\eea
These reduce to the our new boundary conditions  (\ref{newBC}), using the following relations
\bea
\rho=r^2,\qquad  \pi^2T_L^2=\frac{4 G_3 \Delta}{l}, \qquad G_6 = 2\pi^2 l^3 G_3.
\eea
The asymptotic symmetry generators and the central charges then reduce to the ones derived in Section \ref{sec:chargesmetric}.

%\section{Conclusion}

\section*{Acknowledgements}
We are grateful to A.~Castro, T.~Hartman, M.~Henneaux, F.~Larsen , J.~Maldacena  for useful conversations, and especially to
S.~Detournay for his early collaboration. This work was supported by NSF grant 1205550.  W.S. would like to thank Hamburg University and DESY for their hospitality during her visit. W.S. is supported in part by the Harvard Society of Fellows, G.C. is a Research Associate of the Fonds de la Recherche Scientifique F.R.S.-FNRS (Belgium)and A.S. is a Fellow at the Radcliffe Institute for Advanced Study. G.C. also acknowledges the University of Amsterdam and funds from a NWO Vici grant for support during the earliest stages of this project.

\section*{Appendices}
\numberwithin{equation}{section}

\setcounter{section}{0}
\section*{A. Conventions}

We use $r$, $t^\pm = t\pm \phi $ and we take the orientation $\varepsilon^{t\phi r} =\varepsilon^{t\phi}=  - 1$ so that $\varepsilon^{+- r}=\varepsilon^{+-}=+1$, $d^2x =dt \wedge d\phi  =+ \frac{1}{2}dt^- \wedge dt^+$ and $\int_{\mathcal M} d^3 x \p_r = \int_{\partial \mathcal M} d^2 x$. We define  $\epsilon^{\mu\nu\alpha}=\varepsilon^{\mu\nu\alpha}/\sqrt{-g}$. Surface charges on the boundary circle $S$ spanned by the circle $\phi$ can be obtained by
\bea
\int_S k_\xi = \int_{2\pi}^0 \frac{1}{2}k^{\mu\nu}_\xi \varepsilon_{\mu\nu\phi} d\phi = -\int_0^{2\pi} d\phi\; k^{r+}_\xi - \int_0^{2\pi} d\phi\; k^{r-}_\xi\, ,\label{sch}
\eea
where $k_\xi = \frac{1}{2}k^{\mu\nu}_\xi \varepsilon_{\mu\nu\alpha} dx^\alpha$ is the surface charge $1$-form.
%Surface charge densities  on the boundary plane are defined at fixed $t^-$ as $k^{r-}_\xi\, ,$.

We choose a $SL(2,\mathbb R)$ representation such that
 \be \hbox{Tr} (\sigma^{(a)}\sigma^{(b)})=\frac{1}{2}\eta^{(a)(b)}\ee
where
\be \eta_{(a)(b)} =\left( \begin{array}{ccc}
  0&-\half &0     \\
  -\half & 0    &0\\
  0&0&1\\
 \end{array}
\right)_{ab},\ee
which lowers the triad indices. The representation is given by Pauli matrices as
\bea
\sigma^{(-)}_\alpha\,^\beta&=&\left(
\begin{array}{cc}
  0&0      \\
  -1& 0    \\
 \end{array}
\right),\quad \sigma^{(+)}_\alpha\,^\beta=\left(
\begin{array}{cc}
  0&1      \\
  0& 0    \\
 \end{array}
\right),\quad \sigma^{(3)}_\alpha\,^\beta=\half\left(
\begin{array}{cc}
  1&0      \\
  0& -1   \\
 \end{array}
\right).
\eea
The triad and spin connection can be written in spinor notation as
\bea e_{\alpha}^{\;\;\beta}=e_{a}\sigma^{a\;\;\beta}_{\alpha}, \quad \omega_{\alpha}^{\;\;\beta}=\half\epsilon_c\,^{ab}\omega_{ab}\sigma^{c\;\;\beta}_{\alpha}, \eea
where $\omega_{ab}=e_a^{\;\nu} (\p_\mu e_{b\nu}-\Gamma_{\mu\nu}^\rho e_{b\rho})dx^\mu$. Our choice of dreibein leads to
\bea
\epsilon^{(+)(-)(3)}=-2.
\eea

\section*{B. Boundary conditions with $\Delta$ varying}
\label{sec:alt}

In this appendix, we discuss a variant of our boundary conditions where  $\Delta$ is allowed to fluctuate but $\bar P(t^+)$ is periodic. The action which provides a good variational principle is simply $S_0$ because evaluating \eqref{varS} gives
\bea
\delta S_0 = \frac{1}{2\pi}\int dt^+ dt^- \Delta \delta (\p_+ \bar P) = 0.
\eea
Note however that the variation vanishes only globally and not locally. One would therefore expect source terms on the right-hand side of boundary field equations in the reduction of the theory if one repeats the analysis done in Section \ref{CSform}.

The canonical infinitesimal charges associated with the asymptotic generators \eqref{crs} are given for a generic solution in the phase space by
\bea
\sdelta Q_{\xi_R} &=& \frac{1}{2\pi}\int_0^{2\pi} d\phi\; \eps(t^+) \Big(\delta \left(\bar L(t^+)  -\Delta(\p_+\bar P(t^+))^2\right)- \p_+\bar P(t^+)\delta \Delta\Big)\nonumber,\\ \label{ch1b}\\
\sdelta Q_{\eta} &=&  \frac{1}{2\pi}\int_0^{2\pi} d\phi\; \sigma(t^+) (\delta\Delta+\delta \Delta \,\p_+\bar P(t^+) +2\Delta \delta \p_+\bar P(t^+)).\label{ch2b}
\eea
The charges are finite and conserved. One wishes to integrate the  charges to get globally defined quantities on the phase space.
$\sdelta Q_{\xi_R}$ and $\sdelta Q_{\eta}$ are not closed one-forms on phase space and cannot be integrated.  However the  combination
\be \sdelta Q_{\xi_R(\epsilon)} -\sdelta Q_{\eta(\epsilon)} = \frac{1}{2\pi}\int_0^{2\pi} d\phi\; \eps(t^+) \delta \big[\bar L(t^+) -\Delta(1+\p_+\bar P(t^+))^2\big] \ee  is exact. Setting $\epsilon =e^{int^+}$ and integrating we obtain the globally defined charge
\be
\bar \cL_n \equiv Q_{\xi_R(e^{i n t^+})-\eta(e^{i n t^+})} =%\bar{\mathcal{L}}_b+
 \frac{1}{2\pi}\int_0^{2\pi} d\phi e^{i n t^+} (\bar L(t^+) -\Delta(1+\p_+\bar P(t^+))^2).
\label{barLnb}\ee
In order to integrate $\delta Q_\eta$ when $\Delta \neq 0$, we must use a vector field $\eta_n={e^{int^+} \over \sqrt{|\Delta|}}$ normalized to a constant which we take to be $\pm 4G\ell$ depending on whether it is spacelike or timelike. We then have \be
\bar \cJ_n \equiv Q_{\eta((|\Delta|)^{-1/2}e^{i n t^+})} =% \bar{\mathcal{J}}_b+
\frac{\Theta(\Delta)}{\pi}\int_0^{2\pi} d\phi e^{i n t^+} \sqrt{|\Delta|}(1+\p_+\bar P(t^+)),
\ee
where $\Theta(\cdot)$ is the sign function.% and $\bar{\mathcal{J}}_b$ is a background value which is not fixed by the classical analysis.

The Dirac bracket between these generators is given by
\bea
i \{\bar \cL_m,\bar \cL_n\} &=& (m-n)\bar \cL_{m+n}+\frac{c_R}{12}m^3\delta_{m,-n},\\
i\{\bar \cL_m,\bar \cJ_n\}&=&-n \bar \cJ_{m+n},\\
i\{\bar \cJ_m,\bar \cJ_n\} &=& \frac{\tilde k_{KM}}{2} m \delta_{m,-n}.
\eea
The central charge and level of the current algebra are given by
\bea
c_R = \frac{3\ell }{2G},\qquad \tilde k_{KM} = -4\Theta(\Delta) \, .\label{ccb}
\eea

Because $\bar \cJ_0$ is a central element of the algebra it is possible to define generators without the awkward  $\Theta$ function by
\be
\bar \cT_n \equiv \frac{1}{2} \Theta(\bar \cJ_0)\bar \cJ_0\bar \cJ_n=
\frac{1}{\pi}\int_0^{2\pi} d\phi e^{i n t^+} \Delta(1+\p_+\bar P(t^+)),\label{barTnb}
\ee
These obey\footnote{Note that $ k_{KM}$, unlike $\tilde k_{KM}$, has dimensions of action as expected for a central term in a Dirac Bracket algebra persisting in the classical limit. The units are scaled out of $\tilde k_{KM}$ by the field-dependent normalization of the generators.}
\be
[\bar \cT_m,\bar \cT_n] = \frac{ k_{KM}}{2} m \delta_{m,-n}~~~~~~
  k_{KM} = -2\bar \cT_0,  \, .\ee
where $\bar \cT_0=-{\ell \over 8G}$ in the global vacuum and $\bar \cT_0 = 2\Delta$ in the black hole sector. These generators correspond  to the unnormalized vector fields  $\sigma(t^+)\p_-$. Usually a  field-dependent level is inconsistent, but here no problem arises because $\cT_0$ is a central element of the algebra. 

In summary, we defined three distinct phase spaces: $\Delta < 0$, $\Delta = 0$ and $\Delta >0$ since we had to divide by $\sqrt{|\Delta|}$ in order to define the generators. The sector $\Delta < 0$ contains the $AdS_3$ vacuum $\Delta =-k/4$ and its Virasoro-Kac-Moody descendants. The Kac-Moody level is $k=c/6$. The sector $\Delta = 0$ contains one class of extremal BTZ black holes and their Virasoro descendants while the Kac-Moody generators are pure gauge. The sector $\Delta >0$ contains the other class of extremal BTZ black holes, non-extremal BTZ black holes and their Kac-Moody-Virasoro descendants. The Kac-Moody level is negative.

%\bibliography{master2}

\begin{thebibliography}{10}

\bibitem{Brown:1986nw}
J.~D. Brown and M.~Henneaux, ``Central charges in the canonical realization of
  asymptotic symmetries: An example from three-dimensional gravity,'' {\em
  Commun. Math. Phys.} {\bf 104} (1986) 207.

\bibitem{Banados:1992wn}
M.~Ba\~nados, C.~Teitelboim, and J.~Zanelli, ``The {B}lack hole in
  three-dimensional space-time,'' {\em Phys. Rev. Lett.} {\bf 69} (1992)
  1849--1851,
\href{http://www.arXiv.org/abs/hep-th/9204099}{{\tt hep-th/9204099}}.
%%CITATION = HEP-TH 9204099;%%.

\bibitem{Banados:1992gq}
M.~Banados, M.~Henneaux, C.~Teitelboim, and J.~Zanelli, ``Geometry of the (2+1)
  black hole,'' {\em Phys. Rev.} {\bf D48} (1993) 1506--1525,
\href{http://www.arXiv.org/abs/gr-qc/9302012}{{\tt gr-qc/9302012}}.
%%CITATION = GR-QC 9302012;%%.

\bibitem{Saida:1999ec}
H.~Saida and J.~Soda, ``{Statistical entropy of BTZ black hole in higher
  curvature gravity},'' {\em Phys.Lett.} {\bf B471} (2000) 358--366,
\href{http://www.arXiv.org/abs/gr-qc/9909061}{{\tt gr-qc/9909061}}.
%%CITATION = GR-QC/9909061;%%.

\bibitem{Henneaux:2002wm}
M.~Henneaux, C.~Mart\'{\i}nez, R.~Troncoso, and J.~Zanelli, ``Black holes and
  asymptotics of 2+1 gravity coupled to a scalar field,'' {\em Phys. Rev.} {\bf
  D65} (2002) 104007,
\href{http://www.arXiv.org/abs/hep-th/0201170}{{\tt hep-th/0201170}}.
%%CITATION = HEP-TH 0201170;%%.

\bibitem{Hotta:2008yq}
K.~Hotta, Y.~Hyakutake, T.~Kubota, and H.~Tanida, ``Brown-henneaux's canonical
  approach to topologically massive gravity,'' {\em JHEP} {\bf 07} (2008) 066,
\href{http://www.arXiv.org/abs/0805.2005}{{\tt 0805.2005}}.
%%CITATION = 0805.2005;%%.

\bibitem{Compere:2008us}
G.~Comp\`ere and D.~Marolf, ``{Setting the boundary free in AdS/CFT},'' {\em
  Class. Quant. Grav.} {\bf 25} (2008) 195014,
\href{http://www.arXiv.org/abs/0805.1902}{{\tt 0805.1902}}.
%%CITATION = 0805.1902;%%.

\bibitem{Henneaux:2009pw}
M.~Henneaux, C.~Martinez, and R.~Troncoso, ``{Asymptotically anti-de Sitter
  spacetimes in topologically massive gravity},'' {\em Phys. Rev.} {\bf D79}
  (2009) 081502,
\href{http://www.arXiv.org/abs/0901.2874}{{\tt 0901.2874}}.
%%CITATION = 0901.2874;%%.

\bibitem{Liu:2009kc}
Y.~Liu and Y.-W. Sun, ``{Consistent Boundary Conditions for New Massive Gravity
  in $AdS_3$},'' {\em JHEP} {\bf 0905} (2009) 039,
\href{http://www.arXiv.org/abs/0903.2933}{{\tt 0903.2933}}.
%%CITATION = ARXIV:0903.2933;%%.

\bibitem{Henneaux:2010fy}
M.~Henneaux, C.~Martinez, and R.~Troncoso, ``{More on Asymptotically Anti-de
  Sitter Spaces in Topologically Massive Gravity},'' {\em Phys.Rev.} {\bf D82}
  (2010) 064038,
\href{http://www.arXiv.org/abs/1006.0273}{{\tt 1006.0273}}.
%%CITATION = ARXIV:1006.0273;%%.

\bibitem{Maldacena:1998uz}
J.~M. Maldacena, J.~Michelson, and A.~Strominger, ``{Anti-de Sitter
  fragmentation},'' {\em JHEP} {\bf 9902} (1999) 011,
  \href{http://www.arXiv.org/abs/hep-th/9812073}{{\tt hep-th/9812073}}.

\bibitem{Anninos:2008fx}
D.~Anninos, W.~Li, M.~Padi, W.~Song, and A.~Strominger, ``{Warped AdS3 Black
  Holes},'' {\em JHEP} {\bf 03} (2009) 130,
\href{http://www.arXiv.org/abs/0807.3040}{{\tt 0807.3040}}.
%%CITATION = 0807.3040;%%.

\bibitem{Compere:2008cv}
G.~Comp\`ere and S.~Detournay, ``{Semi-classical central charge in
  topologically massive gravity},'' {\em Class. Quant. Grav.} {\bf 26} (2009)
  012001,
\href{http://www.arXiv.org/abs/0808.1911}{{\tt 0808.1911}}.
%%CITATION = 0808.1911;%%.

\bibitem{Guica:2008mu}
M.~Guica, T.~Hartman, W.~Song, and A.~Strominger, ``{The Kerr/CFT
  Correspondence},'' {\em Phys. Rev.} {\bf D80} (2009) 124008,
\href{http://www.arXiv.org/abs/0809.4266}{{\tt 0809.4266}}.
%%CITATION = 0809.4266;%%.

\bibitem{Compere:2009zj}
G.~Comp\`ere and S.~Detournay, ``{Boundary conditions for spacelike and
  timelike warped AdS3 spaces in topologically massive gravity},'' {\em JHEP}
  {\bf 08} (2009) 092,
\href{http://www.arXiv.org/abs/0906.1243}{{\tt 0906.1243}}.
%%CITATION = 0906.1243;%%.

\bibitem{Castro:2009jf}
A.~Castro and F.~Larsen, ``{Near Extremal Kerr Entropy from AdS(2) Quantum
  Gravity},'' {\em JHEP} {\bf 0912} (2009) 037,
  \href{http://www.arXiv.org/abs/0908.1121}{{\tt 0908.1121}}.

\bibitem{Compere:2009qm}
G.~Comp\`ere, S.~de~Buyl, S.~Detournay, and K.~Yoshida, ``{Asymptotic
  symmetries of Schrodinger spacetimes},'' {\em JHEP} {\bf 0910} (2009) 032,
\href{http://www.arXiv.org/abs/0908.1402}{{\tt 0908.1402}}.
%%CITATION = ARXIV:0908.1402;%%.

\bibitem{Anninos:2010pm}
D.~Anninos, G.~Comp\`ere, S.~de~Buyl, S.~Detournay, and M.~Guica, ``{The
  Curious Case of Null Warped Space},'' {\em JHEP} {\bf 1011} (2010) 119,
  \href{http://www.arXiv.org/abs/1005.4072}{{\tt 1005.4072}}.

\bibitem{Guica:2010sw}
M.~Guica, K.~Skenderis, M.~Taylor, and B.~van Rees, ``{Holography for
  Schrodinger backgrounds},''
\href{http://www.arXiv.org/abs/1008.1991}{{\tt 1008.1991}}.
%%CITATION = 1008.1991;%%.

\bibitem{Hofman:2011zj}
D.~M. Hofman and A.~Strominger, ``{Chiral Scale and Conformal Invariance in 2D
  Quantum Field Theory},'' {\em Phys.Rev.Lett.} {\bf 107} (2011) 161601,
  \href{http://www.arXiv.org/abs/1107.2917}{{\tt 1107.2917}}.

\bibitem{ElShowk:2011cm}
S.~El-Showk and M.~Guica, ``{Kerr/CFT, dipole theories and nonrelativistic
  CFTs},'' {\em JHEP} {\bf 1212} (2012) 009,
\href{http://www.arXiv.org/abs/1108.6091}{{\tt 1108.6091}}.
%%CITATION = ARXIV:1108.6091;%%.

\bibitem{Song:2011sr}
W.~Song and A.~Strominger, ``{Warped AdS3/Dipole-CFT Duality},'' {\em JHEP}
  {\bf 1205} (2012) 120,
\href{http://www.arXiv.org/abs/1109.0544}{{\tt 1109.0544}}.
%%CITATION = ARXIV:1109.0544;%%.

\bibitem{Guica:2011ia}
M.~Guica, ``{A Fefferman-Graham-Like Expansion for Null Warped AdS(3)},''
\href{http://www.arXiv.org/abs/1111.6978}{{\tt 1111.6978}}.
%%CITATION = ARXIV:1111.6978;%%.

\bibitem{Azeyanagi:2012zd}
T.~Azeyanagi, D.~M. Hofman, W.~Song, and A.~Strominger, ``{The Spectrum of
  Strings on Warped $AdS_3 \times S^3$},''
\href{http://www.arXiv.org/abs/1207.5050}{{\tt 1207.5050}}.
%%CITATION = ARXIV:1207.5050;%%.

\bibitem{Detournay:2012pc}
S.~Detournay, T.~Hartman, and D.~M. Hofman, ``{Warped Conformal Field
  Theory},''
\href{http://www.arXiv.org/abs/1210.0539}{{\tt 1210.0539}}.
%%CITATION = ARXIV:1210.0539;%%.

\bibitem{Loran:2009cr}
F.~Loran and H.~Soltanpanahi, ``{5D Extremal Rotating Black Holes and CFT
  duals},'' {\em Class.Quant.Grav.} {\bf 26} (2009) 155019,
  \href{http://www.arXiv.org/abs/0901.1595}{{\tt 0901.1595}}.

\bibitem{Azeyanagi:2009wf}
T.~Azeyanagi, G.~Comp\`ere, N.~Ogawa, Y.~Tachikawa, and S.~Terashima,
  ``{Higher-Derivative Corrections to the Asymptotic Virasoro Symmetry of 4d
  Extremal Black Holes},'' {\em Prog. Theor. Phys.} {\bf 122} (2009) 355--384,
\href{http://www.arXiv.org/abs/0903.4176}{{\tt 0903.4176}}.
%%CITATION = 0903.4176;%%.

\bibitem{Balasubramanian:2009bg}
V.~Balasubramanian, J.~de~Boer, M.~M. Sheikh-Jabbari, and J.~Simon, ``{What is
  a chiral 2d CFT? And what does it have to do with extremal black holes?},''
  {\em JHEP} {\bf 02} (2010) 017,
\href{http://www.arXiv.org/abs/0906.3272}{{\tt 0906.3272}}.
%%CITATION = 0906.3272;%%.

\bibitem{Blagojevic:2009ek}
M.~Blagojevic and B.~Cvetkovic, ``{Asymptotic structure of topologically
  massive gravity in spacelike stretched AdS sector},'' {\em JHEP} {\bf 09}
  (2009) 006,
\href{http://www.arXiv.org/abs/0907.0950}{{\tt 0907.0950}}.
%%CITATION = 0907.0950;%%.

\bibitem{Henneaux:2011hv}
M.~Henneaux, C.~Martinez, and R.~Troncoso, ``{Asymptotically warped anti-de
  Sitter spacetimes in topologically massive gravity},'' {\em Phys.Rev.} {\bf
  D84} (2011) 124016,
\href{http://www.arXiv.org/abs/1108.2841}{{\tt 1108.2841}}.
%%CITATION = ARXIV:1108.2841;%%.

\bibitem{css2}
G.~Comp\`ere, W.~Song, and A.~Strominger, ``{Chiral Liouville Gravity},''
  (2013)
\href{http://www.arXiv.org/abs/1303.xxxx}{{\tt 1303.xxxx}}.
%%CITATION = ARXIV:0803.3621;%%.

\bibitem{Coussaert:1995zp}
O.~Coussaert, M.~Henneaux, and P.~van Driel, ``The asymptotic dynamics of
  three-dimensional {E}instein gravity with a negative cosmological constant,''
  {\em Class. Quant. Grav.} {\bf 12} (1995) 2961--2966,
  \href{http://www.arXiv.org/abs/gr-qc/9506019}{{\tt gr-qc/9506019}}.

\bibitem{Detournay:2010rh}
S.~Detournay, D.~Israel, J.~M. Lapan, and M.~Romo, ``{String Theory on Warped
  $AdS_{3}$ and Virasoro Resonances},'' {\em JHEP} {\bf 1101} (2011) 030,
\href{http://www.arXiv.org/abs/1007.2781}{{\tt 1007.2781}}.
%%CITATION = ARXIV:1007.2781;%%.

\bibitem{Skenderis:1999nb}
K.~Skenderis and S.~N. Solodukhin, ``{Quantum effective action from the AdS /
  CFT correspondence},'' {\em Phys.Lett.} {\bf B472} (2000) 316--322,
\href{http://www.arXiv.org/abs/hep-th/9910023}{{\tt hep-th/9910023}}.
%%CITATION = HEP-TH/9910023;%%.

\bibitem{Abbott:1981ff}
L.~F. Abbott and S.~Deser, ``Stability of gravity with a cosmological
  constant,'' {\em Nucl. Phys.} {\bf B195} (1982)
76.
%%CITATION = NUPHA,B195,76;%%.

\bibitem{Barnich:2001jy}
G.~Barnich and F.~Brandt, ``Covariant theory of asymptotic symmetries,
  conservation laws and central charges,'' {\em Nucl. Phys.} {\bf B633} (2002)
  3--82,
\href{http://arXiv.org/abs/hep-th/0111246}{{\tt hep-th/0111246}}.
%%CITATION = HEP-TH 0111246;%%.

\bibitem{Barnich:2007bf}
G.~Barnich and G.~Comp\`ere, ``{Surface charge algebra in gauge theories and
  thermodynamic integrability},'' {\em J. Math. Phys.} {\bf 49} (2008) 042901,
\href{http://www.arXiv.org/abs/0708.2378}{{\tt 0708.2378}}.
%%CITATION = 0708.2378;%%.

\bibitem{Achucarro:1987vz}
A.~Achucarro and P.~K. Townsend, ``{A Chern-Simons Action for Three-Dimensional
  anti-De Sitter Supergravity Theories},'' {\em Phys. Lett.} {\bf B180} (1986)
89.
%%CITATION = PHLTA,B180,89;%%.

\bibitem{Witten:1988hc}
E.~Witten, ``(2+1)-dimensional gravity as an exactly soluble system,'' {\em
  Nucl. Phys.} {\bf B311} (1988)
46.
%%CITATION = NUPHA,B311,46;%%.

\bibitem{Compere:2007vx}
G.~Comp\`ere, ``{Note on the First Law with p-form potentials},'' {\em
  Phys.Rev.} {\bf D75} (2007) 124020,
\href{http://www.arXiv.org/abs/hep-th/0703004}{{\tt hep-th/0703004}}.
%%CITATION = HEP-TH/0703004;%%.

\bibitem{Compere:2009dp}
G.~Comp\`ere, K.~Murata, and T.~Nishioka, ``{Central Charges in Extreme Black
  Hole/CFT Correspondence},'' {\em JHEP} {\bf 0905} (2009) 077,
  \href{http://www.arXiv.org/abs/0902.1001}{{\tt 0902.1001}}.

\end{thebibliography}

\providecommand{\href}[2]{#2}\begingroup\raggedright\endgroup

\end{document}